\documentclass[aps,prl,preprint,showpacs,showkeys,a4paper,superscriptaddress]{revtex4}
\usepackage{graphicx}
\usepackage{amsmath}
\begin{document}
\title{Graphene on the C-terminated SiC (000 $\bar{1}$) surface: An ab initio study }

\author{L.Magaud}
\affiliation{Institut N\'eel, CNRS-UJF, BP 166, 38042 Grenoble Cedex 9 France\\}

\author{F.Hiebel}
\affiliation{Institut N\'eel, CNRS-UJF, BP 166, 38042 Grenoble Cedex 9 France\\}

\author{F.Varchon}
\affiliation{Institut N\'eel, CNRS-UJF, BP 166, 38042 Grenoble Cedex 9 France\\}

\author{P.Mallet}
\affiliation{Institut N\'eel, CNRS-UJF, BP 166, 38042 Grenoble Cedex 9 France\\}

\author{J.-Y.Veuillen}
\affiliation{Institut N\'eel, CNRS-UJF, BP 166, 38042 Grenoble Cedex 9 France\\}

\date{\today}

\begin{abstract}
The atomic and electronic structures of a graphene layer on top of the $(2\times2)$ reconstruction of the SiC (000$\bar{1}$) surface are studied from ab initio calculations. At variance with the (0001) face, no C bufferlayer is found here. Si adatoms passivate the substrate surface so that the very first C layer presents a linear dispersion characteristic of graphene. A small graphene-substrate interaction remains in agreement with scanning tunneling experiments (F.Hiebel et al. {\it Phys. Rev. B} {\bf 78} 153412 (2008)). The stacking geometry has little influence on the interaction which explains the rotational disorder observed on this face.
\end{abstract}


\maketitle

Most of graphene exciting properties such as anomalous quantum Hall effect, high electron mobility, linear dispersion and electronic chirality ~\cite{Novoselov05,Zhang05,Berger06,Bostwick07,Zhou07,Brihuega08} are directly related to the symmetry of the graphene honeycomb lattice. Any coupling to the external world can disturb the lattice or break its symmetry. A substrate that supports graphene without altering its electronic properties  or at least with an interaction as small as possible is then highly desirable. The C-terminated face of the polar SiC $(000\bar{1})$ surface seems to fulfill these conditions, even in a better way than the Si-terminated face does. ~\cite{Berger06,Varchon08,Hiebel08}

In the graphene-on-SiC system, the C layers are produced by sublimating Si atoms
from the polar faces (0001) (Si terminated) or $(000\bar{1})$ (C terminated) of hexagonal SiC polytypes. The annealing takes place at sufficiently high temperature so that the C atoms in excess can rearrange to form  mono or multilayers of graphene. ~\cite{Vanbommel75,Forbeaux,Berger04} The SiC surfaces go through different reconstructions when the temperature is increased. The phases observed by LEED and STM are different for the Si and C face.

On the Si face, the system goes through the $\sqrt{3}\!\times\! \sqrt{3}R30$, $ 6\sqrt{3}\!\times\! 6\sqrt{3}R30$ and then graphene 1x1 symmetry. The first carbon (buffer) layer is in strong interaction with the substrate and passivates  the SiC surface so that the subsequent C planes are only slightly coupled to the substrate. As a consequence they exhibit an electronic structure that is characteristic of graphene ~\cite{Bostwick07,Zhou07,Brihuega08,Varchon_prl07,Varchon_prb08,Mattausch_prl07, Kim_prl08, nous_STM07,Rutter07}. The graphene lattice is aligned with respect to the substrate -with an angle of $30^\circ$- so that in this case a real epitaxy of the graphene layer on the SiC is obtained.

On the C face, the $6\sqrt{3}\!\times\! 6\sqrt{3}R30$ reconstruction is not observed. The SiC bare C-terminated surface exhibits a $3\times3$ reconstruction that can be covered by graphene as reported in  ~\cite{Hiebel08,Vanbommel75,Bernhardt}. When the annealing time or temperature are increased, the underlying $3\times3$ reconstruction appears to evolve into a $(2\!\times\!2)_C$ ~\cite{Hiebel08,Vanbommel75,Forbeaux,Bernhardt,Emtsev_C}. STM data show that the coupling between the first C layer and the substrate is small although not negligible ~\cite{Hiebel08,Emtsev_C}. It is much smaller than on the Si-surface and a linear dispersion is observed on the ARPES spectra ~\cite{Emtsev_C}. Little is known on the decoupling mechanism at the graphene C-terminated SiC interface and this is the object of the present letter. We will first discuss the C-terminated SiC bare surface in the $(2\!\times\!2)_C$ reconstruction and then what happens when it is covered by graphene.

Calculations are performed using the code VASP ~\cite{vasp} that is based on the Density functional theory. We are using the generalized gradient approximation \!\cite{pw} and ultrasoft pseudopotentials \!\cite{uspp}. In all the cells calculated here, the 4H SiC substrate is modelled by a slab that contains 4 SiC bilayers with H saturated DB on the second surface of the slab.
The empty space ranges from 10 to 8 \!\AA. The plane wave basis
cutoff is equal to 211 \!eV. The ultra-soft pseudopotentials have been extensively tested ~\cite{Varchon_prl07,Hass_prl08,Varchon_prb08}. Integration over the Brillouin zone is performed using a 15x15x9 (resp. 6x6x5) grid for the free surface (resp. graphene covered) in the Monckhorst-Pack scheme to ensure
convergence of the Kohn Sham eigenvalues. When graphene is involved, the K point at the corner of the Brillouin zone is included to get a correct description of the Dirac point. Residual forces in the converged structures are smaller than 0.015 eV /\!\AA.

\begin{figure}
\includegraphics[height=9.0cm,angle=-90,clip]{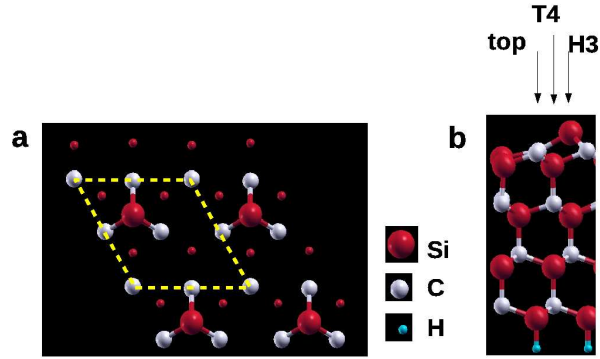}
\caption{(color on line) $(2\!\times\!2)_C$ reconstruction with Si adatoms in H3 position.The unit cell is shown in dashed lines. (a) top view, (b) side view.}
\label{f.1}
\end{figure}

A model ~\cite{Seubert00} based on a silicon adatom has been proposed for the $(2\!\times\!2)_C$ reconstruction of the C-terminated bare SiC surface. It is shown in Fig. ~\ref{f.1}(a). The $2\times2$ cell contains 4 atoms per atomic plane so that the bulk truncated surface has four dangling bonds (DB) -one per surface C atom-. A silicon adatom can take three high symmetry positions on a SiC (111) surface: H3, T4 and top. They are shown in Fig.~\ref{f.1}(b). The top position can be ruled out because it saturates only one DB and indeed, our ab initio calculation shows that the corresponding energy is more than 2 eV above the other. A silicon adatom in H3 or T4 position saturates three DB. We are then left with two DB, one on the Si adatom and the second one on the unsaturated surface C - we will call it the C rest atom in analogy with the DAS model of the Si(111) 7x7 reconstruction ~\cite{DAS}. 

\begin{figure}
\includegraphics[width=8.cm,clip]{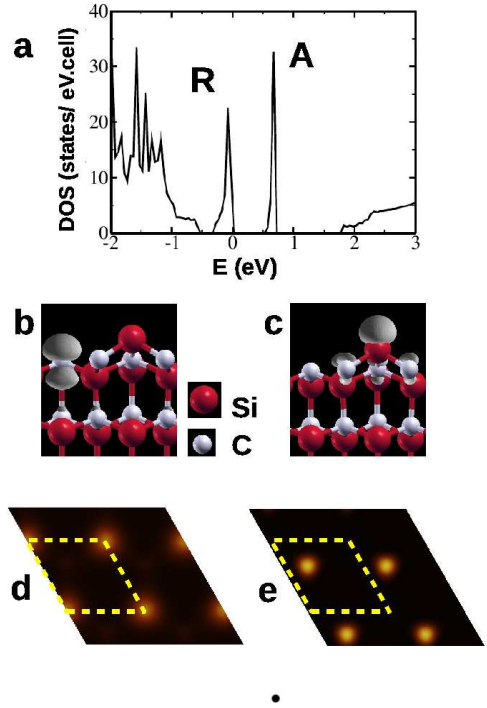}
\caption{(color on line) Electronic structure of the $(2\!\times\!2)_C$ reconstruction of the C-terminated SiC surface. (a) density of states, (b) (resp. (c)) isocontours of $\left|{\psi}\right|^2$ integrated on the energy range of peak R (resp. A), superimposed to the ball and stick model of the surface. (d) (resp. (e)) section of (b) (resp. (c)) just above the adatom. (d) ( resp. (e)) can be compared to a filled (empty) state STM image}
\label{f.2}
\end{figure}

At variance with the Si (111) surface, the H3 site was proposed to be the most stable position here ~\cite{Seubert00}. We find that the T4 position is 1.12 eV higher in energy (see EPAPS). In the H3 configuration, the electronegativity difference between C and Si atoms facilitates a charge transfer from the Si adatom DB to the C restatom DB. As shown in Fig. ~\ref{f.2}(a) this results in a semiconducting surface with two separated narrow bands -named R and A- in the SiC gap.  The Fermi level is at the top of the first band (R). Fig.~\ref{f.2}(b)(d) (resp. (c) and (e)) show the square of the wavefunction modulus integrated over the energy range of peak R (resp. A). The isocontours (b) and (c) demonstrate that peak R is related to the restatom DB that is then filled and that peak A comes from the adatom DB that is empty. The sections (d) and (e) show one protrusion per cell located on the rest atom for filled state (d) and on the Si adatom for empty state (e). They compare very well to STM images ~\cite{Hiebel08,Seubert00,Bernhardt} that show one protrusion per cell at different position in filled and empty states ~\cite{Fanny}. The two DB are found to be either filled or empty and the surface to be passivated. Graphene grown on this face would then be less bound to the substrate than for the Si-terminated case, in agreement with experiments ~\cite{Hiebel08, Emtsev_C}. 

\begin{figure}
\includegraphics[width=9.0cm,clip]{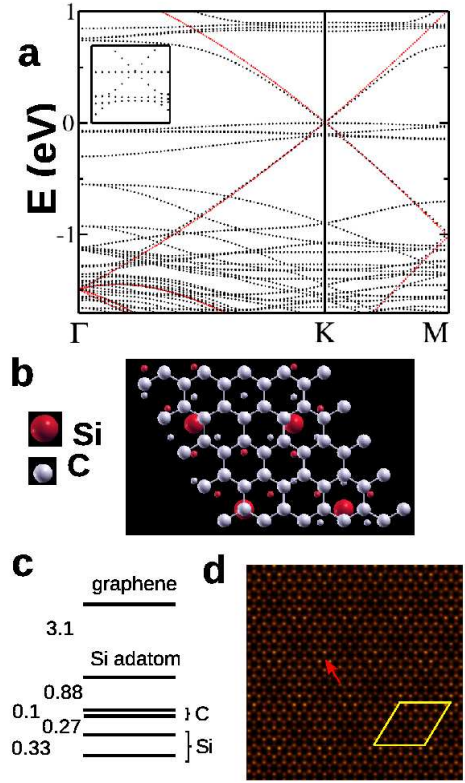}
\caption{(color on line) A $5\times5$ graphene cell on top of a $4\times4$ C-terminated SiC substrate in the $(2\!\times\!2)_C$ reconstruction, in the top configuration.  (a) band structure, the inset shows a zoom in the vicinity of the K point and the dispersion of an isolated graphene layer is given in red (thin dotted lines), (b) top view of the supercell, only the last SiC bilayer, the Si adatoms and the graphene atoms are shown. The size of the atoms is related to their height in the supercell -larger at the surface-. (c) Interlayer distances in \AA.(d) cross section of $\left|{\psi}\right|^2$ integrated from 0.5 to 1.0. The red arrow points to one switched off atom, the supercell is indicated in yellow (white).}
\label{f.3}
\end{figure}

The adatom induces a strong relaxation of the surface in rather good agreement with LEED investigations ~\cite{Seubert00} (see Fig. \ref{f.3} (c)): Its three nearest neighbors C atoms (C-NN) are located on a plane 0.89 \AA\ below the adatom plane. The C rest atom is located 0.11 \AA\ below the C-NNs. The three Si atoms first neighbors of the rest atom are 0.59 \AA\ below the rest atom plane while the remaining Si atom is 0.26 \AA\ below the rest atom plane. The second SiC bilayer is also disturbed, the Si to C height ranges from 0.57 to 0.79 \AA. The third bilayer is almost equivalent to a bulk bilayer.

Graphitisation of the C face is characterised by the appearance of a ring on LEED patterns ~\cite{Hiebel08, Vanbommel75, Forbeaux, Bernhardt,Emtsev_C,Conrad_jpcm08}. It shows that on the $(2\times2)_C$ reconstruction, the graphene or graphitic layers are rotated with respect to each other and to the substrate. It is a crucial property of this face because it has been demonstrated that rotated C layers exhibit an electronic structure similar to the one found for an isolated graphene monolayer ~\cite{Berger06,Hass_prl08,Castroneto,Latil,Pankratov,Sadowski}. This rotational disorder is already observed for the first graphene layer ~\cite{Hiebel08}. It indicates a small coupling of the graphene layer to the substrate. 

Due to the rotational disorder, no well defined common cell is observed on STM images ~\cite{Hiebel08,Fanny}. To build the periodic system that is required in our ab initio study, we used the geometry of the $(2\!\times\!2)_C$ reconstruction and put it in a $4\times4$ supercell -that is with 4 Si adatoms and 4 C rest atoms, see Fig.~\ref{f.3}(b)-. This supercell is nearly commensurate with a graphene $5\times5$ cell without rotation. The mismatch is smaller than 0.2\%. Three different calculations were done (see EPAPS for drawings of configurations (ii) and (iii)): (i) a C atom of graphene on top of one of the Si adatom (top configuration, Fig. ~\ref{f.3}(b)), (ii) a C atom of graphene on top of a Si adatom first neighbours (NN), (iii) a Si adatom below the center of a honeycomb lattice hexagon (center). In this latter case, a graphene atom lies just above a C rest atom. In the supercell, the other graphene / adatom respective positions are shifted with respect to the reference one so that the three remaining adatoms fall in the vicinity of a graphene C atom in cases (i) and (ii) or below the middle of C-C bond in case (iii). These configurations describe the various local environments occuring in systems with rotational disorder. The total energy is not very sensitive to the graphene /SiC relative position and in all cases, we obtain differences that are smaller than 15 meV for the whole supercell. It is a first indication that the position of the graphene layer is not strongly imposed by the substrate. In the following, we focuss on the top case that has the lowest total energy, the results also stand for the other configurations.

The geometry of the SiC $(2\!\times\!2)_C$ reconstruction remains almost unchanged below the graphene layer (Fig. ~\ref{f.3}(c). The mean graphene -  Si adatom distance is equal to 3.1 \AA\. It is much larger than the first C layer-substrate distance on the Si face. The corrugation of the graphene layer remains small -0.07. \AA\ - Because of the Si adatom induced relaxation of the SiC first bilayer, the geometry described here is much more complex than the one proposed for thick samples on the basis of soft X-ray diffraction results. ~\cite{Conrad_jpcm08}

The band structure  is shown in Fig.~\ref{f.3}(a). It  establishes three very important points: 
(i) a linear dispersion is obtained close to the Fermilevel. It means that the first C layer has the electronic structure of a graphene layer in agreement with ARPES ~\cite{Emtsev_C}. This points to a rather weak graphene-substrate coupling in agreement with the large calculated interlayer distance.
(ii) the linear dispersion is alteres by the interaction with interface states and especially with empty states - mainly because the Si adatom DB is closer to the graphene layer than the C restatom one. This highlights the existence of a residual graphene-substrate interaction in agreement with STM data. (iii) The Fermi level falls at the Dirac point.

A light contrast can be seen on the two halves of the graphene $5\times5$ cell. (Fig.~\ref{f.3}(d)) shows a map of the modulus of the square of the wavefunction drawn on a section plane 0.14 \AA\ above the highest atom of the graphene plane. $\left|{\psi}\right|^2$ is integrated on an energy range that goes from 0.5 eV to 1 eV that is in the region where the graphene and the adatom DB interact. Closer to the Fermi level the maps (not shown) shows a honeycomb lattice and for energies below the Fermi level, the graphene contribution is hidden by the restatoms related states. The maximum of the intensity is observed at the middle of the $p_z$ anti-node so that the contrast observed in the two halves of the supercell reflects the small corrugation found in the graphene layer (0.07 \AA\ ). STM images show Moir\'e patterns that can be related to the half cell contrast. The Moir\'e basin observed experimentaly can be larger than the one shown here. This has to be related to the rotation between the graphene and the substrate that can bring more Si adatoms in correspondance with graphene atoms.

On Fig.~\ref{f.3}(d) map, the graphene C atom above the Si adatom -one of them is indicated by a red arrow- is nearly switched off. The position of the section plane does not change the contrast on this atom revealing an electronic effect : the 2D electron gas of graphene interacts with the adatom localized state and the interaction creates a dip in the graphene DOS. "Missing atoms"  have also been observed on STM images and they were found on top of Si adatoms ~\cite{Hiebel08,Fanny}.  The agreement found between the present calculations and STM images of a graphene layer on top of a $(2\!\times\!2)_C$ reconstructed SiC shows that most of the experimental configuration is described by the graphene covered $(2\!\times\!2)_C$ model considered here.

The neutral character of the graphene found in the three calculated configurations is more puzzling since experiments gives a Fermi level 0.2 eV above the Dirac point ~\cite{Emtsev_C,Berger06}. This discrepancy might come from the significant density of defects at the interface observed by STM ~\cite{Hiebel08,Fanny}. Only a very small fraction of electron (0.001) is needed to shift the Fermi level to 0.2 eV above the Dirac point.

At variance with the Si-face where the SiC surface is passivated by a C layer -the bufferlayer- that imposes the orientation of the graphene layers and their stacking sequence, here the SiC surface is passivated by adatoms so that a nearly negligible influence of the graphene-SiC $(2\times2)_C$ stacking is found. Flakes can then grow with different orientations within a layer and also in successive layers.

In conclusion, our ab initio calculations explain why the first C layer on top of a C-terminated SiC surface presents a linear dispersion and how this dispersion is disturbed due to a small but non negligible interaction with the substrate. The  $(2\!\times\!2)_C$ reconstruction  efficiently passivates the SiC surface. Two interface states remain on this surface. One comes from a DB on the Si adatom that is then rather close to the graphene layer but interacts in an energy range away from the Dirac point. The second DB is on the restatom, interaction with the graphene  layer occurs in the vicinity of the Dirac point but remains small because the rest atom is far from the graphene layer. The total energy of the system is not very sensitive to the relative graphene/SiC stacking and explain why a rotational disorder is observed on this face. This gives a clue in the search for graphene on a substrate with a coupling as small as possible : search for passivated surfaces or passivate them on purpose by adding adatoms for instance during or after the growth.

 \subsection*{Acknowledgments}
This work is supported by a computer grant at IDRIS-CNRS; the ACI CIMENT (phynum project); an ANR GRAPHSIC project, a CIBLE07 and a RTRA project. Figures are plotted using Xcrysden.

\end{document}